\documentclass[aps,prl,twocolumn]{revtex4}

\usepackage{graphicx}
\usepackage{amsmath}

\begin{document}

\title{Statistics of temperature fluctuations in an electron system out of equilibrium}

\author{T.~T. Heikkil\"a}
\email[]{Tero.Heikkila@tkk.fi} \affiliation{Low Temperature
Laboratory, Helsinki University of Technology, P.O. Box 5100
FIN-02015 TKK, Finland}

\author{Yuli Nazarov}
\affiliation{Kavli Institute of NanoScience, Delft University of
Technology, 2628 CJ Delft, The Netherlands}

\date{\today}

\begin{abstract}
We study the statistics of the fluctuating electron temperature in a
metallic island coupled to reservoirs via resistive contacts and
driven out of equilibrium by either a temperature or voltage
difference between the reservoirs. The fluctuations of temperature
are well-defined provided that the energy relaxation rate
inside the island exceeds the rate of energy exchange with the
reservoirs. We quantify these fluctuations in the regime beyond the
Gaussian approximation and elucidate their dependence on the nature of
the electronic contacts.
\end{abstract}

\maketitle

The temperature of a given system is well-defined in the case when
the system is coupled to and in equilibrium with a reservoir at
that temperature. Out of equilibrium, the temperature is
determined by a balance of the different heat currents
from/to the system \cite{giazotto06}. However, this applies only to
the average temperature: the heat currents fluctuate giving rise
to temperature fluctuations. Although the equilibrium fluctuations
have been discussed in textbooks \cite{landaulifshitz}, their
existence was still debated around the turn of 1990's
\cite{kitteletal}.

\begin{figure}[h]
\centering
\includegraphics[width=\columnwidth]{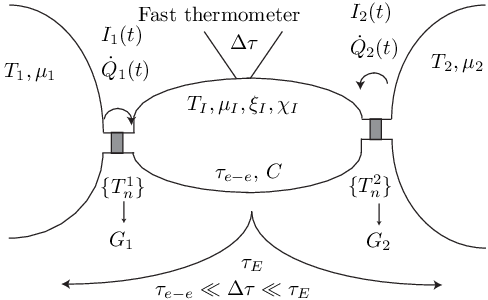}
\caption{Setup and limit considered in this work: a conducting
island is coupled to reservoirs via electrical contacts
characterized by the transmission eigenvalues $\{T_n^\alpha\}$. The
temperature fluctuates on a time scale $\tau_E$ characteristic for
the energy transport through the junctions. We assume the limit
$\tau_{e-e} \ll \tau_E$ where the internal relaxation within the
island is much faster than the energy exchange with reservoirs. In
this limit both the temperature and its fluctuations are well
defined.} \label{fig:tfssetup}
\end{figure}

In this Letter we generalize the concept of temperature fluctuations
to the nonequilibrium case by quantifying their statistics in an
exemplary system: a metal island coupled to two reservoirs (see
Fig.~\ref{fig:tfssetup}). The island can be biased either by a
voltage or temperature difference between the reservoirs. In this
case, the temperature of the electrons is not necessarily well
defined.  The electron-electron scattering inside the island may
however provide an efficient relaxation mechanism to drive the
energy distribution of the electrons towards a Fermi distibution
with a well-defined, but fluctuating temperature
\cite{giazotto06,steinbach96}. Here we assume this {\it
quasiequilibrium} limit where the time scale $\tau_{e-e}$ of
internal relaxation is much smaller than the scale $\tau_{E}$
related to the energy exchange with the reservoirs.

In equilibrium, the only relevant parameters characterizing the
temperature fluctuation statistics (TFS) are the average temperature
$T_a$, fixed by the reservoirs, and the heat capacity $C =\pi^2
k_B^2 T_a/(3\delta_I)$ of the system. The latter is inversely
proportional to the effective level spacing $\delta_I$ on the
island. In terms of these quantities, the probability of the
electrons being at temperature $T_e$ reads
\cite{landaulifshitz,normalizationnote}
\begin{equation}\label{eq:eqflucs}
P_{\rm eq}(T_e) \propto \exp\left[-\frac{C
(T_e-T_a)^2}{k_BT_a^2}\right] = \exp\left[-\frac{\pi^2 k_B
(T_e-T_a)^2}{3T_a \delta_I}\right],
\end{equation}
corresponding to Boltzmann distribution of the total energy of the
island. The probability has a Gaussian form even for large
deviations from $T_a$, apart from that the probability naturally
vanishes for $T_e < 0$. From this distribution we can for example
infer the variance, $\langle (\Delta T_I)^2 \rangle = k_BT^2/C$. As
we show below, the scale of the probability log, $\ln P \sim
T_a/\delta_I$ is the same for the nonequilibrium case while its
dependence on $(T_e/T_a)$ is essentially different.

To generalize the concept of temperature fluctuations to
the nonequilibrium case we examine the probability that the
temperature of the island measured within a time interval $\tau_0
\dots \tau_0+\Delta \tau$ and averaged over the interval, equals $T_e$:
\begin{equation}\label{eq:average0}
\begin{split}
&P(T_e)=\left\langle \delta_I\left(\frac{1}{\Delta
\tau}\int_{\tau_0}^{\tau_0+\Delta \tau} T_I(t)dt -T_e\right)
\right\rangle\\ &= \left\langle \int \frac{dk}{2\pi}
\exp\left[ik\left(\int_{\tau_0}^{\tau_0+\Delta \tau} (T_I(t)-T_e)dt
\right)\right]\right\rangle.
\end{split}
\end{equation}
The average $\langle \cdot \rangle$ is over the nonequilibrium state
of the system. The latter is evaluated using an extension of the Keldysh
technique \cite{keldysh65} where the fluctuations of charge and heat
are associated with two counting fields, $\chi$ and $\xi$, respectively
\cite{nazarov02,kindermann04,pilgram04}. The technique
allows one to evaluate the full statistics of current
fluctuations both for charge \cite{nazarov02} and heat current
\cite{kindermann04} in an arbitrary multiterminal system. In terms
of the fluctuating temperature and chemical potential of the island, $T_I(t)$
and $\mu_I(t)$, and the associated counting fields $\xi_I(t)$ and
$\chi_I(t)$ the average in Eq.~\eqref{eq:average0} is presented in
the form
\begin{equation}\label{eq:average1}
\begin{split}
P(T_e) \propto & \int {\cal D}\xi_I(t) {\cal D}T_I(t) {\cal
D}\chi_I(t) {\cal D}\mu_I(t) dk \\&\times \exp\left\{-{\cal
A}+ik\left[\int_{\tau_0}^{\tau_0+\Delta \tau} dt (T_I(t) - T_e)
\right]\right\}.
\end{split}
\end{equation}
Here ${\cal A}={\cal A}[\xi_I(t),T_I(t),\chi_I(t),\mu_I(t)]$ is the
Keldysh action of the system. The counting fields $\xi_I(t)$ and
$\chi_I(t)$ enter as Lagrange multipliers that ensure the
conservation of charge and energy \cite{pilgram04}.

The Keldysh action consists of two types of terms, ${\cal A}=\int dt
(S_I(t)+S_{\rm c}(t))$, with $S_I(t)=Q_I \dot{\chi}_I  + E_I
\dot{\xi}_I$ describing the storage of charge and heat on the island
and $S_{\rm c}$ describing the contacts to the reservoirs. Here
$Q_I=C_c \mu_I$ is the charge on the island, $E_I=C(T_I) T_I/2 + C_c
\mu_I^2/2$ gives the total electron energy of the island and
$C_c$ is the electrical capacitance of the island. For
the electrical contacts, the action can be expressed in terms of the
Keldysh Green's functions as \cite{snyman08} (we set $\hbar=e=k_B=1$
for intermediate results)
\begin{equation}
S_{c,el} = \frac{1}{2} \sum_\alpha \sum_{n\in \alpha} {\rm Tr}\ {\rm
ln}\left[1+T_n^\alpha
\frac{\{\check{G}_{\alpha},\check{G}_I\}-2}{4}\right].
\end{equation}
The sums run over the lead and channel indices $\alpha$ and $n$. All
products are convolutions over the inner time variables. The trace
is taken over the Keldysh indices and the action is evaluated with
equal outer times. This action is a functional of the Keldysh
Green's functions $\check{G}_{\alpha}$ and $\check{G}_I$ of the
reservoirs and the island, respectively. It also depends on the
transmission eigenvalues $\{T_n^\alpha\}$, characterizing each
contact. The counting fields enter the action by the gauge
transformation of Green's function \cite{kindermann04}
\begin{equation}
\check{G}(t,t')=e^{-\frac{1}{2}(\chi_I-i\xi_I(t)
\partial_t)\check \tau_3}
\check{G}_0(t,t') e^{\frac{1}{2}(\chi_I+i\xi_I(t')
\partial_{t'})\check \tau_3}.
\end{equation}
where the Keldysh Green's function reads
\begin{equation}
\check{G}_0(t,t')=\int \frac{d\epsilon}{2\pi} e^{-i \epsilon(t-t')}
\begin{pmatrix}
1-2 f(\epsilon) & 2 f(\epsilon)\\
2-2f(\epsilon) & -1 + 2 f(\epsilon)
\end{pmatrix}.
\end{equation}
For quasiequilibrium $f(\epsilon)=\{\exp[(\epsilon-\mu)/T]+1\}^{-1}$
is a Fermi distribution. In what follows, we assume the fields
$\xi(t), T(t)$ to vary slowly at the time scale $T^{-1}$, in which case
we can approximate $-i\xi(t) \partial_t \mapsto \xi(t) \epsilon$.

The saddle point of the total action at $\chi=\xi=0$ yields the
balance equations for charge and energy. Assuming that the
electrical contacts dominate the energy transport, we get
\begin{subequations}
\begin{align}
\frac{\partial Q_I}{\partial t}&=C_c \partial_t \mu_I = \sum_\alpha
{\rm Tr}\check \tau_3 \sum_n T_n^\alpha \frac{[\check
G_\alpha,\check G_I]}{4+T_n^\alpha (\{\check G_I,\check
G_\alpha\}-2)}\\
\begin{split}
\frac{\partial E_I}{\partial t}&=C \partial_t T_I \\&=\sum_\alpha
{\rm Tr}(\epsilon-\mu_I)\check \tau_3 \sum_n T_n^\alpha
\frac{[\check G_\alpha,\check G_I]}{4+T_n^\alpha (\{\check
G_I,\check G_\alpha\}-2)}.
\end{split}
\label{eq:energycontinuity}
\end{align}
\end{subequations}
The right-hand sides are sums of the charge and heat currents,
respectively, flowing through the contacts $\alpha$
\cite{nazarov99}.

The time scale for the charge transport is given by $\tau_c=C_c/G$,
with $G= \sum_\alpha \sum_n T_n^\alpha/(2 \pi)$. This is typically
much smaller than the corresponding time scale for heat transport,
$\tau_E=C_h/G_{\rm th}$, where $G_{\rm th}=\pi^2 G T/3$. We assume
that the measurement takes place between these time scales, $\tau_c
\ll \Delta \tau \ll \tau_E$. In this limit the potential and its
counting field $\mu_I$ and $\chi_I$ follow adiabatically the
$T_I(t)$ and $\xi_I(t)$ and there is no charge accumulation on the
island. As a result, we can neglect the charge capacitance $C_c$
concentrating on the zero-frequency limit of charge transport.

To determine the probability, we evaluate the path integral in
Eq.~\eqref{eq:average1} in the saddle-point approximation. There are
four saddle-point equations,
\begin{subequations}
\begin{align}
\partial_{\chi_I} S_c&=0, \quad \partial_{\mu_I} S_c=0\label{eq:chargesaddlepoint}\\
\frac{\pi^2}{6}\frac{\dot{\xi}_I}{\delta_I}&=-\partial_{T_I^2} S_c-\frac{ikM_b(t;\tau_0,\Delta \tau)}{2T_I}\\
\frac{\pi^2}{6}\frac{\dot{T}_I^2}{\delta_I}&=\partial_{\xi_I} S_c.
\end{align}
\end{subequations}
Here $M_b(t)=1$ inside the measurement interval
$(\tau_0,\tau_0+\Delta\tau)$ and zero otherwise. Equations
\eqref{eq:chargesaddlepoint} express the chemical potential and
charge counting field in terms of instant values of temperature
$T_I$ and energy counting field $\xi_I$, $\mu_I=\mu_I(\xi_I,T_I)$
$\chi_I=\chi_I(\xi_I,T_I)$. The third and fourth equations give the
evolution of these variables. It is crucial for our analysis that
these equations are of {\it Hamilton} form,  $\xi_I$ and $T_I^2$
being conjugate variables, the total connector action $S_c$ being an
integral of motion. Boundary conditions at $t \pm \infty$ correspond
to most probable configuration $T_e=T_a$. This implies $S_c=0$ at
trajectories of interest.

\begin{figure}[h]
\centering
\includegraphics[width=\columnwidth]{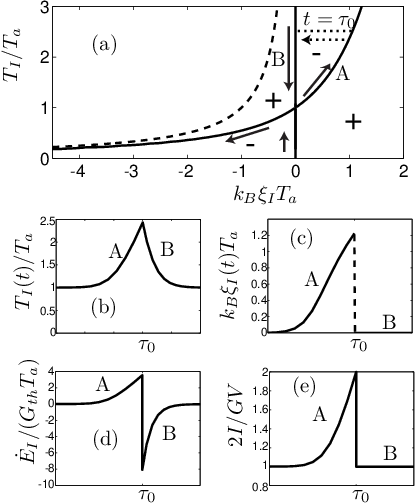}
\caption{Time line of a huge fluctuation. The measurement is made at
$t \approx \tau_0$ with result $T_e=2.5 T_a$. For $t<\tau_0$, the
temperature follows the "anti-relaxation" branch A, whereas after
the measurement, it relaxes as predicted by a "classical" equation.
(a) Contour plot of $S_c$, evaluated for the case of two equal
tunnel junctions with $T_1=T_2=0$ and bias voltage $V$
($T_a=\sqrt{3}eV/(2\pi k_B))$. The zeroes  of $S_c$ are given by
crossing curves; the dashed line indicates the boundary of the
domain where $\xi$ is defined ($\xi>-1/(k_B T)$). The sign of $S_c$
in the different regions is also indicated. The arrows give the
direction of the flow for the saddle-point solutions
$\{\xi_I^S(t),T_I^S(t)\}$. (b) and (c) show the time dependence of
the fluctuation for $T_I(t)$ and $\xi_I(t)$, respectively. The heat
current into the island corresponding to this fluctuation is plotted
in (d), and (e) shows the charge current flowing through the
island.} \label{fig:effaction}
\end{figure}
The zeros of $S_c$ in $\xi_I-T_I$ plane are concentrated in two
branches that cross at the equilibrium point $\xi_I=0,T_e=T_a$ (Fig.
\ref{fig:effaction} a). Branch B ($\xi=0$) corresponds to the usual
"classical" relaxation to the equilibrium point from either higher
or lower temperatures. Branch A corresponds to "anti-relaxation":
the trajectories following the curve quickly depart from equilibrium
to either higher or lower temperatures. The solution of the
saddle-point equations follows A before the measurement and B after
the measurement (Figs. \ref{fig:effaction} b-e).

Since $S_c=0$, the only contribution to path integral
\eqref{eq:average1} comes from the island term $C \dot \xi_I T_I^2$
and is evaluated as
\begin{equation}\label{eq:average2}
P(T_e) = \exp\left[\frac{\pi^2}{3\delta_I}\int \dot{\xi} T^2
dt\right] = \exp\left[-\frac{2\pi^2}{3\delta_I} \int_{T_a}^{T_e} T
\xi_I^S(T) dT\right].
\end{equation}
Thus, in order to find $P(T_e)$, we only need a function
$\xi_I^S(T_I)$ satisfying $S_c(\xi_I^S(T_I),T_I)=0$ at branch A.

The connector action can generally be written in the form
\begin{equation}
\begin{split}
S_c=&\sum_\alpha \sum_{n\in \alpha} \int \frac{d\epsilon}{2\pi} \ln
\{1+T_n^\alpha [f_I(1-f_\alpha)\\&\times(e^{-\chi_I-\xi_I
\epsilon}-1)+f_\alpha(1-f_I)(e^{\chi_I+\xi_I \epsilon}-1)]\},
\end{split}
\end{equation}
with
$f_{\alpha/I}=\{\exp[(\epsilon-\mu_{\alpha/I})/T_{\alpha/I}]+1\}^{-1}$.

To prove the validity of the method for
the equilibrium case, let us set all the chemical
potentials to 0 and all the reservoir temperatures
 to $T_a$. This implies $\mu_I=\chi_I=0$.
Using the fact that for a Fermi
function $f=-e^{\epsilon/T}(1-f)$, we observe that
 $S_c=0$ regardless of contact properties
 provided $\xi_I=\xi_I^S(T_I)=1/T_L-1/T_I$.
Substituting this to Eq.~\eqref{eq:average2} reproduces
the equilibrium distribution, Eq.~\eqref{eq:eqflucs}.

\begin{figure}[h]
\centering
\includegraphics[width=\columnwidth]{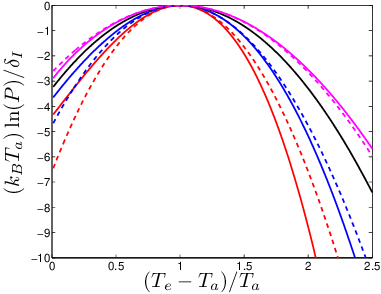}
\caption{(Color online.) Logarithm of TFS probability $P(T_e)$ in a
few example cases. Solid lines from top to bottom: temperature bias
with symmetric tunneling contacts, $T_a=T_1/\sqrt{2}$, $T_2=0$
(magenta); Gaussian equilibrium fluctuations (black), nonequilibrium
fluctuations with $T_a=\sqrt{3}|eV|/(2\pi k_B)$, $T_1=T_2=0$ for
symmetric tunneling and ballistic contacts (blue and red lines,
respectively). The dashed lines are Gaussian fits to small
fluctuations $(T_e-T_a)\ll T_a$, described by the heat current noise
$S_Q$ at $T_e \approx T_a$.} \label{fig:prob}
\end{figure}

Out of equlibrium, the further analytical progress
 can be made
in the case when the connectors are ballistic, $T_n\equiv 1$. Such a
situation can be realized in a chaotic cavity connected to terminals
via open quantum point contacts. The connector action reads
\cite{pilgram04},
\begin{equation}
\begin{split}
S_c=&\sum_\alpha \frac{G_\alpha}{2} \bigg[\frac{2 \mu_\alpha
\chi_I+T_\alpha \chi_I^2 + \left[\pi^2
T_\alpha^2/3+\mu_\alpha^2\right]\xi_I}{1-T_\alpha \xi_I}\\&-\frac{2
\mu_I \chi_I-T_I \chi_I^2 + \left[\pi^2
T_I^2/3+\mu_I^2\right]\xi_I}{1+T_I \xi_I}\bigg].
\end{split}
\end{equation}
Let us first assume two reservoirs with $T_1=T_2\equiv T_L$. In this
case the general saddle-point solution for the potential follows
from Kirchoff law: $\mu_I=(g \mu_1+ \mu_2)/(1+g)$ with $g\equiv
G_L/G_R$. For the charge counting field we get $\chi_I=-\mu_I \xi$.
The most probable temperature $T_a$ is given by $T_a^2=T_L^2+3g
(\mu_1-\mu_2)^2/[\pi^2(1+g)^2]$, and function $\xi_I^S(T_I)$ is
expressed as
\begin{equation}
\xi_I^S=\frac{T_I^2-T_a^2}{T_I(T_LT_I+T_a^2)}.
\end{equation}
Substituting this to Eq.~\eqref{eq:average2} yields for the full
probability
\begin{equation}\label{eq:chaoticcavitygenprob}
\begin{split}
-\ln P_{\rm ball}= \frac{\pi^2 k_B}{3\delta_I T_L^3}
\bigg[&T_L(T_e-T_a)((T_e+T_a)T_L-2
T_a^2)\\&+2T_a^2(T_a^2-T_L^2)\ln\left(\frac{T_a^2+T_e T_L}{T_a^2+T_a
T_L}\right)\bigg].
\end{split}
\end{equation}
In the strong nonequilibrium limit $V\equiv(\mu_1-\mu_2) \gg T_L$,
i.e., $T_a \gg T_L$ this reduces to
\begin{equation}
P_{\rm ball} \propto \exp\left\{-\frac{2\pi^2 k_B}{3\delta_I}
\frac{(T_e+2 T_a)(T_e-T_a)^2}{3 T_a^2}\right\}.
\end{equation}
The logarithm of this probability is plotted as the lowermost line
in Fig.~\ref{fig:prob}.

If the island is biased by temperature difference,
$T_1 \equiv T_L \gg T_2$, $V=0$,
the probability obeys the same Eq.~\eqref{eq:chaoticcavitygenprob} with
$T_a^2=g T_1^2/(1+g)$.

For general contacts, the connector action and its
saddle-point trajectories have to be calculated numerically. For {\em tunnel}
contacts, the
full probability distribution
 is plotted in two regimes in
Fig.~\ref{fig:prob}. The distribution takes values between the
ballistic and equilibrium cases. Let us understand this concentrating
on Gaussian regime and inspecting
the variance of the temperature fluctuations for various contacts.
This variance is related to
the zero-frequency heat current noise $S_{\dot Q}$ via
\begin{equation}
2 G_{\rm th} C\langle \delta T^2 \rangle = S_{\dot Q}
=\partial_\xi^2 S_c|_{\xi \rightarrow 0}.
\end{equation}
In equilibrium, $S^{(eq)}_{\dot Q} = 2 G_{\rm th} T^2$ by virtue of
the fluctuation-dissipation theorem. For an island with equal
ballistic contacts driven far from equilibrium, $V \gg T_L$,
$S_{\dot Q}^{\rm bal} =\sqrt{3} GV^3/(8\pi) =G_{\rm th}(T_a) T_a^2$,
i.e., only half of $S^{(eq)}_{\dot Q}$. The reduction manifests
vanishing temperature of the reservoirs. Most generally, for
contacts of any nature, the heat current noise reads
\begin{equation}
S_{\dot Q}/S^{(eq)}_{\dot Q} = \frac{1}{2} + a_Q \sum_\alpha
F_\alpha,
\end{equation}
where $F_\alpha=\sum_n T_n^\alpha (1-T_n^\alpha)/\sum_n T_n^\alpha$
is the Fano factor for a contact $\alpha$, $a_Q \approx 0.112$ being
a numerical factor.
For two tunnel contacts we hence
obtain $S_{\dot Q}^{\rm tun} \approx 0.723 S^{(eq)}_{\dot Q}$,
a value between the ballistic and equilibrium values. For contacts of any
type,  the variation of temperature fluctuations is between the
ballistic and tunneling values.

For rare fluctuations of temperature, $|T_e-T_a| \simeq T_a$, the
probability distribution is essentially non-Gaussian in contrast to
the equilibrium case. The skewness of the distribution is  negative
in the case of voltage driving: low-temperature fluctuations ($T_e <
T_a$) are preferred to the high-temperature ones ($T_e > T_a$). In
contrast, biasing with a temperature difference (uppermost curve in
Fig.~\ref{fig:prob}) favours high-temperature fluctuations.

The non-Gaussian features of the temperature fluctuations can be
accessed at best in islands with a large level spacing, that is
smaller than the average temperature say, by an order of magnitude.
Many-electron quantum dots with spacing up to 0.1 K$/k_B$ seem
natural candidates for the measurement of the phenomenon. The most
natural way to detect the rare fluctuations is through a threshold
detector \cite{tobiska04}, which produces a response only for
temperatures exceeding or going under a certain threshold value.
Besides the direct measurement of temperature, one can use the
correlation of fluctuations. For example,
Fig.~\ref{fig:effaction}(e) shows that the fluctuation of the
temperature also causes a fluctuation in the charge current.
Observing the latter may thus yield information about the former.

To conclude, we have evaluated non-equilibrium temperature
fluctuations of an example system beyond the Gaussian regime. The
method makes use of saddle-point trajectories and allows to describe
electric contacts of arbitrary transparency.

We thank M. Laakso for useful comments on the manuscript. This work
was supported by the Academy of Finland and the Finnish Cultural
Foundation. TTH acknowledges the hospitality of the Delft University
of Technology, where this work was initiated.


\begin{thebibliography}{14}
\bibitem{normalizationnote} In this Letter, we evaluate the
probabilities with exponential accuracy. The pre-exponential factor
at $T_e \approx T_a$ can be determined by the normalization.
\end{thebibliography}
\end{document}